\documentclass[aps,prl,
groupedaddress,amssymb,
showpacs,nofootinbib]{revtex4}

\usepackage{amsmath,amssymb}

\usepackage[usenames]{color}

\usepackage{graphicx}
\usepackage{mathptmx}
\usepackage{bbm}
\usepackage{bm}
\usepackage{times}

\newcommand{\beq}{\begin{eqnarray}}
\newcommand{\eeq}{\end{eqnarray}}

\makeatother

\begin{document}

\onecolumngrid


\title{The universal coefficient of the exact correlator of a large-$N$ matrix field theory}

\author{Eytan \surname{Katzav}$^{a}$}

\email{ eytan.katzav@mail.huji.ac.il}

\author{\;Peter \surname{Orland}$^{bcd}$}

\email{orland@nbi.dk}

\affiliation{a. The Racah Institute of Physics, The Hebrew University of Jerusalem, Jerusalem 91904, Israel}

\affiliation{b. The Niels Bohr Institute, The Niels Bohr International Academy,
Blegdamsvej 17, DK-2100, Copenhagen {\O}, Denmark}

\affiliation{c. Baruch College, The 
City University of New York, 17 Lexington Avenue, 
New 
York, NY 10010, U.S.A. }

\affiliation{d. The Graduate School and University Center, The City University of New York, 365 Fifth Avenue,
New York, NY 10016, U.S.A.}

\begin{abstract}

Exact expressions have been proposed for correlation functions of
the large-$N$ (planar) limit of the 
$(1+1)$-dimensional ${\rm SU}(N)\times {\rm SU}(N)$ principal chiral sigma model. These were obtained
with the form-factor bootstrap. The short-distance form of the two-point function of the scaling field $\Phi(x)$, was 
found to be $N^{-1}\langle {\rm Tr}\,\Phi(0)^{\dagger} \Phi(x)\rangle=C_{2}\ln^{2}mx$, where $m$ is the mass gap, in
agreement with the perturbative renormalization group. Here we point out that the universal coefficient $C_{2}$, is proportional to
the mean first-passage time of a L\'{e}vy flight in one dimension. This observation enables 
us to calculate $C_{2}=1/16\pi$.

\end{abstract}

\maketitle

\section{I. Introduction}
\setcounter{equation}{0}
\renewcommand{\theequation}{1.\arabic{equation}}

The main problem of quantum chromodynamics is to understand quark confinement and the mass gap. An analogous
problem with similar features, {\em e.g.}, asymptotic freedom and non-trivial anomalous dimensions, is the principal
chiral sigma model (PCSM) of a matrix field $U(x) \in {\rm SU}(N)$, $N\ge 2$, where $x^{0}$ and 
$x^{1}$ are the time and space coordinates, respectively. The action is
\begin{eqnarray}
S=\frac{N}{2g_{0}^{2}}\int d^{2}x \;\eta^{\mu\nu}\;{\rm Tr}\,\partial_{\mu}U(x)^{\dagger}\partial_{\nu}
U(x),
\label{action}
\end{eqnarray}
where $\mu, \nu=0,1$, $\eta^{00}=1$, $\eta^{11}=-1$, $\eta^{01}=\eta^{10}=0$, and $g_{0}$ is the coupling. This action does not change under the global transformation 
$U(x)\rightarrow V_{L}U(x)V_{R}$,  for two matrices $V_{L}, \,V_{R}\in {\rm SU}(N)$. The scaling or renormalized field operator $\Phi(x)$ is an average of $U(x)$ over a region of size $b$, where $\Lambda^{-1}<b\ll m^{-1}$, where $\Lambda$ is the ultraviolet cutoff and $m$ is the mass of the fundamental excitation. The normalization of $\Phi$ is
determined by 
\begin{eqnarray}
\langle 0\vert \Phi(0)_{b_{0} a_{0}}\vert P,\theta, a_{1}, b_{1}\rangle
=N^{-1/2}\delta_{a_{0} a_{1}}\delta_{b_{0} b_{1}}, \label{norm}
\end{eqnarray}
where the ket on the right is a one particle (hence the symbol $P$) state, with rapidity $\theta$ (that is, with momentum components $p_{0}=m\cosh\theta$, $p_{1}=m\sinh\theta$) and we implicitly sum over 
left and right colors
$a_{1}$ and $b_{1}$, respectively.

In the bootstrap approach for some two-dimensional field theories, the exact S matrix \cite{ZZ} and form factors \cite{FF} can
be found heuristically, using the powerful
property of integrability. On the other hand, the bootstrap begins from one's expectations
about the mass spectrum, rather than proving these expectations. In particular, one must assume the 
existence
of a mass gap $m$. In our opinion, the ultimate goal of
the bootstrap should be to {\em reconstruct} the Lagrangian or Hamiltonian formulation of the 
quantum field theory (we will say more about this towards the end of the paper). This would provide
a proof of the existence of the mass gap in the latter formulations. A more modest 
step forward \cite{log-squared}, was to show that if 
$N\rightarrow\infty$ \cite{'t}, $m\vert x\vert \ll 1$, the bootstrap expression of the two-point function of
the scaling field $\Phi(x)$, in Euclidean space, has the behavior
\beq
N^{-1}\langle 0\vert {\rm Tr}\,{\mathcal T}\;\Phi(0)^{\dagger} \Phi(x)\vert 0\rangle \simeq C_{2}\ln^{2}(m\vert x\vert), \label{starting-point}
\eeq
where $\mathcal T$ denotes time ordering. This result was obtained from the exact expression for the Wightman (non-time ordered) two-point function
in Minkowski space \cite{Large-N-FF1}, \cite{Large-N-FF2}:
\begin{eqnarray}
{\mathcal W}(x)=N^{-1}\langle 0\vert {\rm Tr}\,\Phi(0)^{\dagger} \Phi(x)\vert 0\rangle
=\int_{-\infty}^{\infty} \frac{d\theta_{1}}{4\pi} e^{{\rm i}p_{1}\cdot x}+ \frac{1}{4\pi}\sum_{l=1}^{\infty} \int_{-\infty}^{\infty} d\theta_{1}\cdots \int_{-\infty}^{\infty} d\theta_{2l+1}
e^{{\rm i}\sum_{j=1}^{2l+1}p_{j}\cdot x}\;
\prod_{j=1}^{2l}\frac{1}{(\theta_{j}-\theta_{j+1})^{2}+\pi^{2}}
\;,
\label{series}
\end{eqnarray}
where $\theta_{j}$ are rapidities and $p_{j}=m(\cosh\theta_{j}, \sinh\theta_{j})$ are the corresponding momentum vectors, for $j=1,\dots 2l+1$. 

In this paper, we will show how to evaluate the coefficient $C_{2}$ in (\ref{starting-point}). For 
pedagogical completeness, we will briefly review the 
derivation of (\ref{series}) in Section II, and how this series was used to find
(\ref{starting-point}), in Section III.

Standard saddle-point large-$N$ methods fail for the PCSM. This is related to the fact that the Feynman diagrams in the large-$N$ limit are planar 
\cite{'t}, instead of
linear. We emphasize that (\ref{starting-point}) is a significant departure from 
$n\rightarrow \infty$ results for simpler isovector
quantum field theories, {\em e.g.}, O($n$) nonlinear sigma models or ${\mathbb C}{\mathbb P}$($n-1$) models. In particular, Eq. (\ref{starting-point}) is not the correlation function of a free field, although a free {\em master} field
does exist \cite{Large-N-FF1}.

The result (\ref{starting-point}) is in perfect agreement with the perturbative renormalization group applied to
the action (\ref{action}). We give a brief summary here (for a more complete discussion, see References \cite{RCV}). Consider
the regularized Euclidean correlation function (obtained after a Wick rotation, $x^{0}\rightarrow {\rm i}x^{0}$) is
$G(\vert x \vert, \Lambda)=N^{-1}\left \langle 0\vert \;{\mathcal T}\;{\rm Tr}\;\Phi(x) \Phi(0)^{\dagger} \vert 0\right\rangle$,
defined with an ultraviolet cut-off $\Lambda$. The ultraviolet behavior of the correlation function may be found from the renormalization group equations:
\begin{eqnarray}
\frac{\partial\ln G(R, \Lambda)}{\partial\ln \Lambda}= \gamma(g_{0}^{2})=\gamma_{\,1}g_{0}^{2}+\cdots\;,\;\;\;
\frac{\partial g_{0}^{2}(\Lambda)}{\partial \ln \Lambda}=\beta(g_{0}^{2})=-\beta_{1}g_{0}^{4}+\cdots, \label{RGE}
\end{eqnarray}
The coefficients of the anomalous dimension $\gamma(g_{0}^{2})$ and the beta function 
$\beta(g_{0}^{2})$ are $\gamma_{\;1}=(N^{2}-1)/(2\pi N^{2})$ and
$\beta_{1}=1/(4\pi)$. For large $\Lambda$, the dimensionless quantity $G(R,\Lambda)$ becomes a function of the product of the two variables $G(R\Lambda)$. Integrating (\ref{RGE})
yields the leading behavior
\begin{eqnarray}
G(R,\Lambda) \sim C[\ln (R\Lambda)]^{\gamma_{\,1}/\beta_{1}}\;. \label{asympt}
\end{eqnarray}
The power of the logarithm is $\gamma_{\,1}/\beta_{1}=2-2/N^{2}$, which becomes $2$ in the limit of infinite $N$.

Note that $C_{2}$ in Eq. (\ref{starting-point}) is a universal quantity, because the normalization of $\Phi(x)$ is set by (\ref{norm}). We will show that $C_{2}=1/16\pi$. In fact, this quantity has already been evaluated in context of
L\'{e}vy flights \cite{BuldEtAl}; it is proportional to the mean-first passage time in one dimension.

In the next section we briefly review the form factors and correlation functions of the scaling field. Then we explain 
how this expression leads to (\ref{starting-point}) and present an expression for $C_{2}$ in terms of the spectrum
of the square-root of the one-dimensional Laplacian $\Delta^{1/2}=\sqrt{-d^{2}/du^{2}}$, with $u\in [-1,1]$, in Section III. In Section IV, we determine the value of 
$C_{2}$. We conclude with a few remarks in Section V.

\section{II. Form factors and correlation functions of the Principal Chiral Sigma Model}
\setcounter{equation}{0}
\renewcommand{\theequation}{2.\arabic{equation}}

The expression (\ref{series}) 
was found from the form factors of the scaling field in the 
large-$N$ limit \cite{Large-N-FF1}, \cite{Large-N-FF2}. These form factors satisfy a set of axioms proposed by
Smirnov \cite{FF} (motivated by the Lehman-Symanzik-Zimmerman formulation of field theory \cite{LSZ}), formulated with the S matrix of the PCSM (for finite $N$) discussed in 
References \cite{Smatrix}. Form factors and correlation functions of other local operators in the PCSM may be found in References
\cite{ACC1}. Some of these results have been extended to a finite volume in Reference \cite{ACC2}. 

The S matrix of the elementary excitations of the principal chiral model 
\cite{Smatrix}, depends upon the incoming rapidities
$\theta_{1}$ and 
$\theta_{2}$ (as discussed in the introduction, the momentum vectors are $(p_{j})_{0}=m\cosh\theta_{j}$,
$(p_{j})_{1}=m\sinh \theta_{j}$),
outgoing rapidities
$\theta_{1}^{\prime}$ and $\theta_{2}^{\prime}$
and rapidity difference $\theta=\vert \theta_{12}\vert =\vert \theta_{1}-\theta_{2}\vert $. In the limit of  large $N$ we assume that 
$m$ is fixed, as 
$N\rightarrow \infty$ (all available evidence indicates that this is 
the standard 't Hooft limit). The
excitations which survive in the large-$N$ limit are elementary particles and elementary
antiparticles. The $1/N$-expansion of the 
two-particle S matrix 
\beq
S_{PP}(\theta)^{c_{2}d_{2};c_{1}d_{1}}_{a_{1}b_{1};a_{2}b_{2}}
=\left[ 1+O(1/N^{2})\right]
\left[\delta^{c_{2}}_{a_{2}}\delta^{d_{2}}_{b_{2}}\delta^{c_{1}}_{a_{1}} \delta^{d_{1}}_{b_{1}}
-\frac{2\pi {\rm i}}{N\theta}\left(
\delta^{c_{2}}_{a_{1}}\delta^{d_{2}}_{b_{2}}\delta^{c_{1}}_{a_{2}} \delta^{d_{1}}_{b_{1}}
+\delta^{c_{2}}_{a_{2}}\delta^{d_{2}}_{b_{1}}\delta^{c_{1}}_{a_{1}} \delta^{d_{1}}_{b_{2}}
\right)
-\frac{4\pi^{2}}{N^{2}\theta^{2}}
\delta^{c_{2}}_{a_{1}}\delta^{d_{2}}_{b_{1}}\delta^{c_{1}}_{a_{2}} \delta^{d_{1}}_{b_{2}}
\right]. \label{expanded-s-matrix}
\eeq
The generalized S matrix is defined
by replacing
$\theta=\vert\theta_{12}\vert$ with
$\theta=\theta_{12}$ in (\ref{expanded-s-matrix}). The 
generalization is necessary to analytically continue rapidities into the complex
plane. The S matrix of one particle and one antiparticle $S_{PA}(\theta)$ is obtained by crossing
(\ref{expanded-s-matrix}) from the $s$-channel to the $t$-channel:
\beq
&S_{PA}\!\!\!&\!\!\!(\theta)^{d_{2}c_{2};c_{1}d_{1}}_{a_{1}b_{1};b_{2}a_{2}}=\left[ 1+O(1/N^{2})\right] \nonumber \\
&\times& \left[
\delta^{d_{2}}_{b_{2}}\delta^{c_{2}}_{a_{2}}\delta^{c_{1}}_{a_{1}}\delta^{d_{1}}_{b_{1}}
-\frac{2\pi{\rm i}}{N{\hat \theta}}
\!\left(\! \delta_{a_{1}a_{2}}\delta^{c_{1}c_{2}}\delta^{d_{2}}_{b_{2}}\delta^{d_{1}}_{b_{1}}
+\delta^{c_{2}}_{a_{2}}\delta^{c_{1}}_{a_{1}}\delta_{b_{1}b_{2}}\delta^{d_{1}d_{2}}
\! \right)\! -\frac{4\pi^{2}}{N^{2}{\hat \theta}^{2}}
\delta_{a_{1}a_{2}}\delta^{c_{1}c_{2}}\delta_{b_{1}b_{2}}\delta^{d_{1}d_{2}}
  \right], \label{crossed-S}
\eeq 
where ${\hat \theta}=\pi {\rm i}-\theta$ is the crossed rapidity difference. 

For a set of particles with labels $1$, $2,$, etc., we denote a particle's rapidity 
$\theta_{j}$, left color $a_{j}$ and right color $b_{j}$ by $P_{j}=\{P, \theta_{j}, a_{j}, b_{j}\}$. For a set of antiparticles with labels $1$, $2,$, etc., we denote an antiparticle's rapidity 
$\theta_{j}$, right color $b_{j}$ and left color $a_{j}$ by $A_{j}=\{P, \theta_{j}, b_{j}, a_{j}\}$ (the reversal of color 
indices is a convention). A multiparticle in-state may be written as 
\beq
\vert \!\!\!&\!\!\!P\!\!\!\!\!&\!\!\!\!\!,\theta_{1}, a_{1}, b_{1};\; P,\theta_{2}, a_{2}, b_{2};\; \cdots; \;P,\theta_{k}, a_{k}, b_{k}; \;
A, \theta_{k+1}, b_{k+1}, a_{k+1}; \; A, \theta_{k+2}, b_{k+2}, a_{k+2};\;\cdots ;\;A, \theta_{k+j}, b_{k+j}, a_{j+1}
\rangle_{\rm in}  \nonumber \\
&=&\vert P_{1};P_{2};\cdots P_{k}; A_{k+1}; A_{k+2};\cdots A_{k+j}
\rangle_{\rm in}. \nonumber
\eeq

By ``form factors", we mean matrix elements of local operators. Using Smirnov's axioms \cite{FF}, form factors of $\Phi(x)$, consistent 
with the S matrix (\ref{expanded-s-matrix}), (\ref{crossed-S}) can be found:
\beq
\!\!\!\!\!\!\!\langle 0\vert \Phi(0)_{b_{0}a_{0}}\vert P_{1};P_{2};\cdots P_{M}; A_{M+1}; A_{M+2};\cdots A_{2M-1}
\rangle_{\rm in}
=
\frac{\sqrt N}{N^{M}} \!\!\sum_{\sigma,\tau\in S_{M}}
F_{\sigma \tau}(\theta_{1},\theta_{2},\dots,\theta_{2M-1})
\!\prod_{j=0}^{M-1}\delta_{a_{j}\;a_{\sigma(j)+M}} 
\delta_{b_{j}\;b_{\tau(j)+M}} ,
\label{M-FF}
\eeq
where the leading part in the $1/N$-expansion of the function $F_{\sigma\tau}=F^{0}_{\sigma\tau}+O(1/N)$ is
\beq
F^{0}_{\sigma \tau}(\theta_{1},\theta_{2},\dots,\theta_{2M-1})
=
\frac{ (-4\pi)^{M-1}K_{\sigma \tau} }{\prod_{j=1}^{M-1} 
[\theta_{j}-\theta_{\sigma(j)+M}+\pi{\rm i}][\theta_{j}-\theta_{\tau(j)+M}+
\pi{\rm i}]},  \label{MFF}
\eeq
where 
\beq
K_{\sigma \tau}=\left\{ \begin{array}{cc} 1\;,& \;\sigma(j)\neq \tau(j), \;{\rm for \;all} \;j
\\
0\;,& \!\!\!\!\!\!\!\!\!\!\!\!\!\!\!\!\!\!\!\!\!\!\!\!\!\!\!\!\!\!\!\!\!\!\!\!{\rm otherwise} 
\end{array} \right.\;\;. \label{Kdef}
\eeq    

We note that (\ref{norm}) agrees with (\ref{M-FF}), (\ref{MFF}) and (\ref{Kdef}), for $M=1$. The case of $M=2$ was solved in Ref. \cite{Large-N-FF1}, while the general case was solved in Ref. \cite{Large-N-FF2}.

The Wightman function is found from the completeness relation:
\beq
{\mathcal W}(x)\; =\;\frac{1}{N}\sum_{a_{0},b_{0}}\sum_{X}\;\langle 0 \vert \Phi(0)_{b_{0}a_{0}} \vert X\rangle_{\rm in}
\;\;_{\rm in}\langle X \vert \;\Phi(0)_{b_{0}a_{0}}^{*} \;\vert 0\rangle\; e^{{\rm i}p_{X}\cdot x} 
\;=\;\frac{1}{N}\sum_{a_{0},b_{0}}\sum_{X}\,
\left\vert \langle 0 \vert \Phi(0)_{b_{0}a_{0}} \vert X\rangle_{\rm in}\right\vert^{2}\,
\;e^{{\rm i}p_{X}\cdot x} , \label{completeness}
\eeq
where $X$ denotes an arbitrary choice of particles, momenta and colors and where $p_{X}$ is the 
momentum eigenvalue of the state $\vert X\rangle$. Substitution of (\ref{M-FF}), (\ref{MFF}) and (\ref{Kdef}) into the completeness relation yields (\ref{series}).

For increasingly large separation $x$, states $\vert X\rangle_{\rm in}$ with many excitations in (\ref{completeness}) 
contribute minimally to ${\mathcal W}(x)$. Therefore, the leading large-distance behavior is 
exponential decay, as expected for a massive theory.  Evaluating correlation functions for small $x$ requires keeping
all the terms of (\ref{completeness}), which is considerably more subtle \cite{log-squared}.

\section{III. Short-distance behavior and the fractional Laplacian}
\setcounter{equation}{0}
\renewcommand{\theequation}{3.\arabic{equation}}

The expression for the Wightman function (\ref{series}) can be studied at short distances, by a method similar to that of Ref. \cite{CMYZ} for the Ising model, using that model's exact form factors \cite{Ising}. We Wick-rotate the time variable to Euclidean space, setting $x^{1}=0$ and
replacing $x^{0}$ by ${\rm i}R$, $R>0$. The phases in (\ref{series}), change via 
$\exp({{\rm i}p_{j}\cdot x}) \rightarrow \exp(-mR\cosh \theta_{j})$. We define $L=\ln \frac{1}{mR}$.
As $mR$ becomes small, $\exp(-mR\cosh \theta_{j})$ becomes approximately the characteristic
function of $(-L,L)$, equal to unity for $-L<\theta<L$ and zero everywhere else. The characteristic
function appears the same way in the Feynman-Wilson gas \cite{FW}. The 
short-distance Euclidean two-point function is now
\begin{eqnarray}
G(mR)
=\frac{L}{2\pi} + \frac{L}{4\pi}\sum_{l=1}^{\infty} \int_{-1}^{1} du_{1}\cdots \int_{-1}^{1} du_{2l+1}\;
\prod_{j=1}^{2l}\frac{1}{L[(u_{j}-u_{j+1})^{2}+(\pi/L)^{2}]}
\;,
\label{Gseries1}
\end{eqnarray}
where $\theta_{j}=Lu_{j}$. 

The terms of  (\ref{Gseries1}) are related to the fractional-power-Laplace operator $\Delta^{1/2}=\sqrt{-d^{2}/du^{2}}$ \cite{review}. This operator acts on a function $f(u)$, vanishing for $u\notin (b,c)$, by \cite{review}
\begin{eqnarray}
{\Delta}^{1/2}f(u)=\frac{1}{\pi}\;{\rm P}\;\int_{b}^{c}du^{\prime}\;\frac{f(u^{\prime})-f(u)}{(u^{\prime}-u)^{2}}\;, \label{FL}
\end{eqnarray}
where P denotes the principal value. We set $b=-1$, $c=1$. The operator $\Delta^{1/2}$ has an infinite
set of discrete eigenvalues $\lambda_{n}$, of the eigenfunctions $\varphi_{n}(u)$, $\Delta^{1/2}\varphi_{n}=\lambda_{n}\varphi_{n}$, $n=1,2,\dots$, with $0<\lambda_{1}<\lambda_{2}<\cdots$, with
$\varphi_{n}(\pm 1)=0$. Here is the relation: for
$u, u^{\prime}\in (-1,1)$, we define the operator $H(L)$ by
\begin{eqnarray}
\frac{1}{L[(u-u^{\prime})^{2}+(\pi/L)^{2}]}
=
\langle u^{\prime} \vert e^{-\frac{\pi}{L}H(L)}
\vert u \rangle. \label{TM}
\end{eqnarray}
By (\ref{FL}), (\ref{TM}) and a straightforward calculation, we find that $H(L)$ is an approximation to $\Delta^{1/2}$, {\em i.e.}, $H(L)={\Delta}^{1/2}+O(1/L)$, with spectrum
\begin{eqnarray}
H(L)\varphi_{n}(u,L)=\lambda_{n}(L)\varphi_{n}(u,L), \; \int_{-1}^{1}du\,\vert\phi_{n}(u,L)\vert^{2}=1,\;\lambda_{n}(L)=\lambda_{n}+O(1/L), \;\varphi_{n}(u,L)=\varphi_{n}(u)
+O(1/L)\;.
\label{pertspec}
\end{eqnarray}

Summing 
over $l$ in
Eq. (\ref{Gseries1}) yields, from (\ref{pertspec}),
\begin{eqnarray}
G(mR)=\frac{L}{4\pi} \int_{-1}^{1}du^{\prime} \int_{-1}^{1} du \; 
\langle \,u^{\prime}\; \vert \;\frac{1}{1-e^{-2\pi H(L)/L}}\; \vert \,\,u \,\rangle 
=\frac{L}{4\pi}\sum_{n=1}^{\infty} \left\vert  \int_{-1}^{1}du\; \varphi_{n}(u,L)      \right\vert^{2} \frac{1}{1-e^{-2\pi\lambda_{n}/L+O(1/L^{2})}}. \label{resum}
\end{eqnarray}
Expanding (\ref{resum}) in powers of $1/L$ (we are expanding 
within the sum, which requires justification. See Ref. 
\cite{log-squared} for a more careful discussion), we find
\begin{eqnarray}
G(mR)
\;=\;\frac{L^{2}}{8\pi^{2}}\sum_{n=1}^{\infty} \left\vert  \int_{-1}^{1}du\; \varphi_{n}(u)      \right\vert^{2} 
\lambda_{n}^{-1}+O(L). \nonumber
\end{eqnarray}
This is (\ref{starting-point}) with the universal coefficient identified as
\begin{eqnarray}
C_{2}=\frac{1}{8\pi^{2}}\sum_{n=1}^{\infty}\left\vert  \int_{-1}^{1}du\; \varphi_{n}(u)      \right\vert^{2} \lambda_{n}^{-1}.\label{C_2}
\end{eqnarray}

\section{IV. Evaluation of $C_{2}$}
\setcounter{equation}{0}
\renewcommand{\theequation}{4.\arabic{equation}}

This section is the heart of this paper. An 
expression which is proportional to the 
right-hand-side of (\ref{C_2}) was evaluated in Reference 
\cite{BuldEtAl}. The result is
\begin{eqnarray}
\sum_{n=1}^{\infty}\left\vert  \int_{-1}^{1}du\; \varphi_{n}(u)  \right\vert^{2} \lambda_{n}^{-1}=\int_{-1}^{1}du\sqrt{1-u^{2}}=\frac{\pi}{2},
\label{1pt}
\end{eqnarray}
which is the mean first-passage time of a L\'{e}vy flight in one dimension, calculated via the corresponding continuum theory, namely the anomalous Laplace equation. Since the discussion in Reference 
\cite{BuldEtAl} may not be easily accessible to readers working in quantum field theory, we present a short derivation
of (\ref{1pt}) below.

We write the square root of the one-dimensional Laplacian (\ref{FL}) as (with $b=-1$, $c=1$)
\begin{eqnarray}
\Delta^{1/2}\;f(u)=\frac{1}{2\pi}\int_{-1}^{1} du^{\prime} \left[ \frac{1}{(u-u^{\prime}-{\rm i}\epsilon)^{2}}
+\frac{1}{(u-u^{\prime}+{\rm i}\epsilon)^{2}}
\right] f(u^{\prime}). \nonumber
\end{eqnarray}
For a function $f(u)$, satisfying the Dirichlet boundary conditions $f(b)=f(c)=0$, this may be integrated by parts to
\begin{eqnarray}
\Delta^{1/2}\;f(u)=-\frac{1}{\pi}\int_{-1}^{1} du^{\prime}\;  \frac{d f(u^{\prime})}{du^{\prime}}\;
{\rm P}\frac{1}{u^{\prime}-u}, \label{practicaldef}
\end{eqnarray}
where P denotes the principal value. Detailed properties of the eigenfunctions $\phi_{n}(u)$ and eigenvalues 
$\lambda_{n}$ can be found in References \cite{spectrum}.

Let us define the function $C(u)$ by 
\begin{eqnarray}
C(u)=\sum_{n=1}^{\infty} \lambda_{n}^{-1}\int_{-1}^{1}du^{\prime} \phi_{n}(u^{\prime}) \phi_{n}(u),
\label{C of u}
\end{eqnarray}
and note that 
\begin{eqnarray}
C_{2}=\frac{1}{8\pi^{2}}\int_{-1}^{1}du \;C(u). 
\end{eqnarray}
This function satisfies
\begin{eqnarray}
\Delta^{1/2}\;C(u)=1, \label{int eq for C}
\end{eqnarray}
by completeness. Now the spectrum of 
$\Delta^{1/2}$
in a finite interval is strictly positive. If there were two square-normalizable solutions to (\ref{int eq for C}), their difference would be a square-normalizable function annihilated
by this positive operator; but this is impossible unless the difference vanishes. Therefore, the square-normalizable solution 
$C(u)$ to (\ref{int eq for C}) is 
unique, and must
be equal to (\ref{C of u}). After presenting the solution to
Eqs. (\ref{practicaldef}) and (\ref{int eq for C}) for $C(u)$, we will integrate to find $C_{2}$ \cite{BuldEtAl}. 

A solution of the integral equation
\begin{equation}
\int_{-1}^{1}du^{\prime}\; \frac{dC(u^{\prime})}{du^{\prime}}\; {\rm P} \frac{1}{(u^{\prime}-u)^{\alpha}}=-1, \;\;\alpha>0,
\label{general eq}
\end{equation}
which is square-normalizable, is
\begin{eqnarray}
C(u)=\frac{1}{\pi\alpha}(1-u^{2})^{\alpha/2}. 
\end{eqnarray}
This is easily checked: the derivative has two branch points at $\pm 1$. Taking the branch cut on
the real axis from $-1$ to $1$, straightforward complex integration yields (\ref{general eq}). Specializing to $\alpha=1$, as in Eq. (\ref{practicaldef}), we obtain the solution to Eq. (\ref{int eq for C}). Upon integrating $C(u)$ we obtain (\ref{1pt}) and
\begin{eqnarray}
C_{2}=\frac{1}{8\pi^{2}}\frac{\pi}{2}=\frac{1}{16\pi},
\end{eqnarray}
which is the result we claimed for the universal coefficient.

\section{V. Remarks}
\setcounter{equation}{0}
\renewcommand{\theequation}{5.\arabic{equation}}

Let us summarize the result of Ref. \cite{log-squared} and this paper. We have found that the correlation function
of the scalar field has the Euclidean asymptotic behavior:
\beq
N^{-1}\langle 0\vert {\rm Tr}\,{\mathcal T}\;\Phi(0)^{\dagger} \Phi(x)\vert 0\rangle \;\simeq \;
\left\{   \begin{array}{c}
\int \frac{d^{2}p}{(2\pi)^{2}}\frac{\exp({\rm i}p\cdot x)}{p^{2}+m^{2}},\; \;\;\vert x\vert \gg m^{-1}
\\    \\
\;\; \frac{1}{16\pi}\ln^{2}(m\vert x\vert),\; \;\;\;\; \vert x\vert \ll m^{-1} 
\end{array}
\right.\;. \label{summary}
\eeq
The normalization of the expressions on the right-hand side is completely determined by (\ref{norm}). The form for large separation $x$, is the Wick rotation of the first term of (\ref{series}). Perhaps the normalization of the short-distance form can be checked with a
lattice Monte-Carlo simulation at relatively large $N$, as has been done for the O($3$) nonlinear sigma model \cite{BNPWW}.

The agreement with the perturbative renormalization group is encouraging, but it is desirable to have a convincing demonstration that the canonical
and bootstrap definitions of the PCSM are the same. We next discuss how it may be possible to determine the regularized Lagrangian directly from the bootstrap, using the trace anomaly.

Although the stress-energy-momentum tensor $T_{\mu\nu}$ of the classical field theory has a vanishing trace, 
$T^{\mu}_{\;\;\mu}=0$, this property is broken in the quantum theory. The trace anomaly for PCSM, with
a point-splitting cut-off $R=(R^{0}, R^{1})$,
\beq
T^{\mu}_{\;\;\mu}&=& \frac{N}{2} \,\frac{dg(R)^{-2}}{d\ln R}
\left[\;{\rm Tr}\;\partial_{\alpha}U^{\dagger}\partial_{\beta}U\vert_{R}
-\langle0\vert\;{\rm Tr}\;\partial_{\alpha}U^{\dagger}\partial_{\beta}U\vert_{R}\vert 0\rangle\right] \nonumber\\
&=&-\frac{N}{2g(R)^{4}} \beta(g)\left[ \; {\rm Tr}\;j^{\mu}(x+R/2)j_{\mu}(x-R/2)
-\langle0\vert\; {\rm Tr}\;j^{\mu}(x+R/2)j_{\mu}(x-R/2)\vert 0\rangle\right],  \label{trace-anomaly}
\eeq
where $\beta(g)$ is the beta function, as before and $j_{\mu}={\rm i}(\partial_{\mu}U)U^{\dagger}$ is a current. Notice
that the right-hand side is proportional to the original Lagrangian; this is a general feature of the trace anomaly. 

Now form factors of both the stress-energy-momentum tensor and current are known \cite{ACC1}. Matrix 
elements of the right-hand side of (\ref{trace-anomaly}) can be calculated. These can be compared with matrix elements of the left-hand 
side. This can be done by working out the operator product expansion of currents, which should have the form (in Minkowski space):
\beq\frac{1}{N}\;{\rm Tr}\;j^{\mu}(R/2)\;j_{\mu}(-R/2) \simeq \frac{1}{128\pi^2}\frac{1}{R^{\alpha}R_{\alpha}}
-2g(R)^{4} \beta(g)^{-1} T^{\mu}_{\;\;\mu}(0)+\cdots. \label{OPE}
\eeq
The first term on the right-hand side of (\ref{OPE}) is the vacuum expectation 
value in (\ref{trace-anomaly}). One of us (P.O.) has calculated this term (not yet published) with
methods similar to those of Reference \cite{log-squared} and of this paper. The global symmetry 
implies that no logarithms can appear in
this term, but, in the context of our bootstrap 
method, it seems miraculous that they do not. By considering matrix elements of 
(\ref{trace-anomaly}) between one-particle states, the second term of (\ref{OPE}) may be determined. 

We hope that the surprising connection between integrable quantum field theory and anomalous diffusion will yield further insights.

\begin{acknowledgments}

P.O. thanks Jan Ambj{\o}rn for discussions. P.O. is supported in part by a grant from the PSC-CUNY.
P.O.'s visit to the Niels Bohr Institute is supported by the ERC-Advanced grant 291092, ``Exploring the Quantum Universe" (EQU).

\end{acknowledgments}

\end{document}